\documentclass[twocolumn,longbibliography,amsmath,amssymb,superscriptaddress]{revtex4-1}
\usepackage[paperwidth=210mm,paperheight=279mm,hmargin=1.5cm,vmargin=2cm]{geometry}
\usepackage{amsmath} 
\usepackage{here}

\usepackage{verbatim} 

\usepackage{graphicx} 
\usepackage{color} 

\usepackage{bm}
\usepackage{subfigure} 
\usepackage{hyperref} 
\usepackage[normalem]{ulem} 
\usepackage{siunitx} 

\begin{document}
\title{Magnetic control of orientational order and intrinsic hydrodynamic instability in bacterial turbulence} 
\author{Kazusa Beppu}
\affiliation{Department of Applied Physics, Aalto University School of Science, Puumiehenkuja 2, Espoo, 02150, Finland}
\author{Jaakko V. I. Timonen}
\affiliation{Department of Applied Physics, Aalto University School of Science, Puumiehenkuja 2, Espoo, 02150, Finland}

\begin{abstract}
\textbf{Highly concentrated active agents tend to exhibit turbulent flows, reminiscent of classical hydrodynamic turbulence, which has attracted considerable attention lately. Controlling the so-called active turbulence has long been a challenge, and the influence of external fields on such chaotic self-organization remains largely unexplored. Here we report on active turbulence of \textit{Bacillus subtilis} bacteria controlled by a uniform magnetic field via a magnetizable medium based on magnetic nanoparticles. The rod-shaped bacteria act as non-magnetic voids in the otherwise magnetic medium, allowing magnetic torques to be generated on their bodies. This leads to an externally controllable nematic alignment constraint that further controls bacterial turbulence into a nematic state. The nematic orientational ordering in the direction parallel to the magnetic field is accompanied by transverse flows owing to active stress by dipole pushers, which induce undulation of the nematic state. Remarkably, the typical length of the undulation is almost independent of the magnetic field strength. Our theoretical model based on the hydrodynamic equations for suspensions of self-propelled particles predicts the intrinsic length scale of hydrodynamic instability independent of the magnetic field. Our findings suggest that magnetic torques are a powerful approach for controlling both individual agents and their collective states in active systems.}
\end{abstract}
\maketitle

Collections of autonomous motile elements which convert energy locally into mechanical motion, the so-called active matter, tend to display a rich variety of collective motion and self-organization\cite{Vicsek,ramaswamy,marchetti}. Intriguingly, such active systems often exhibit self-sustained turbulent flows consisting of transient vortices and jets, named active turbulence, reminiscent of classical hydrodynamic turbulence\cite{Alert}. Paradigmatic examples range from biological systems, e.g., swimming bacteria\cite{Dombrowski,Wensink,Li,Peng,aranson,Patteson}, eukaryotic cells\cite{Mercader,Lin}, and sperm cells\cite{Creppy}, to non-biological systems, e.g., synthetic colloidal particles\cite{Nishiguchi}. Although in the regime of low Reynolds numbers, active turbulence is self-organized by local energy injection from their constituent elements. Due to its ubiquity, active turbulence has attracted great attention during the last two decades. However, controlling the active turbulence has turned out to be challenging. While several efforts have focused on controlling the turbulence through boundary conditions via static geometric walls\cite{Wioland1,Wioland2,Wioland3,Beppu1,dogic,Nishiguchi2,hardouin,Beppu2}, the highly desired ability to control all active particles in the bulk of the liquid using, e.g., external fields remains largely unexplored and unachieved. 

\begin{figure*}[t]
\begin{center}
\includegraphics[width=180mm]{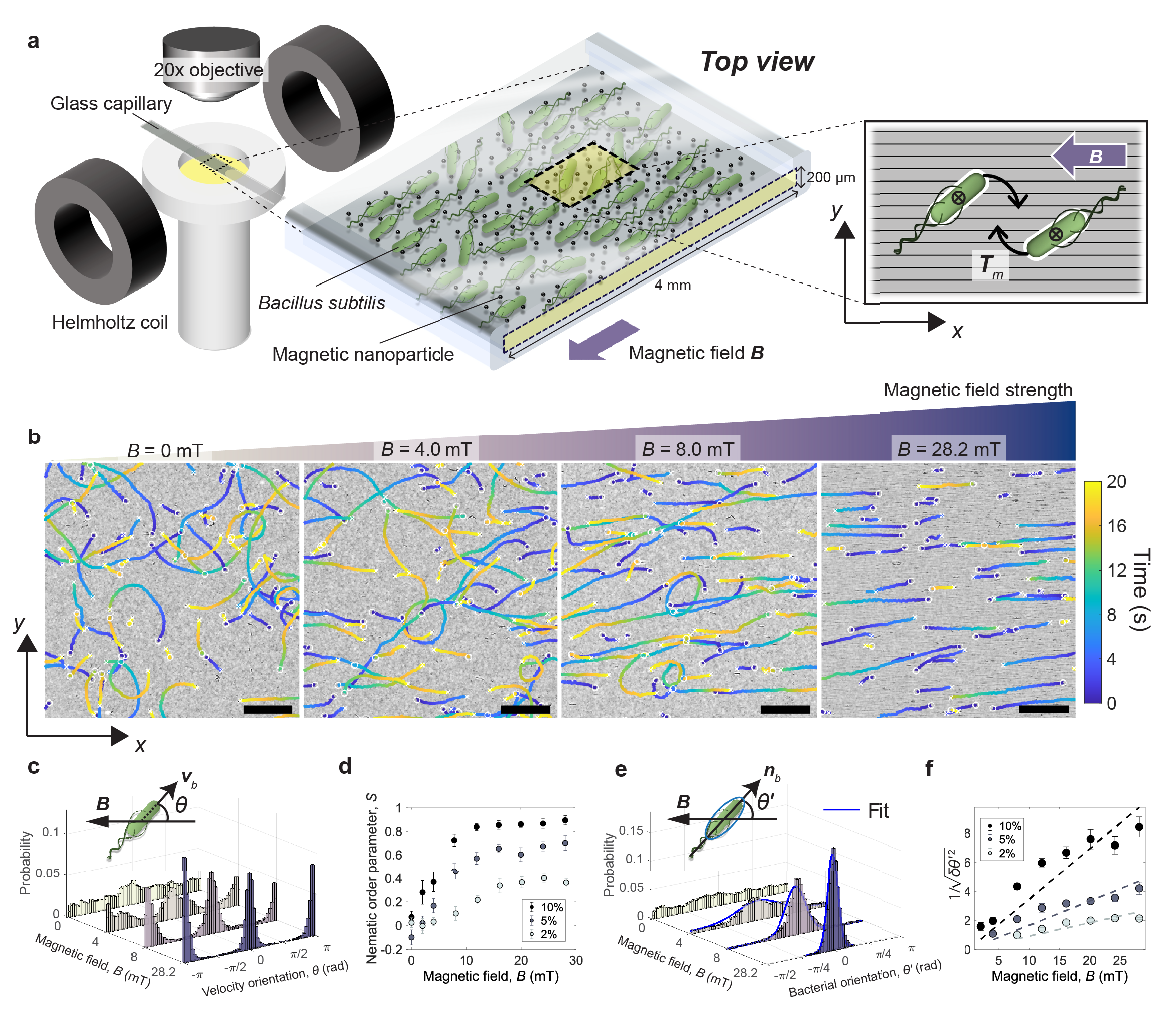}
\end{center}
\caption{\textbf{Magnetic control of swimming direction of non-magnetic bacteria.} (a) A schematic illustration of the experimental setup, showing a rectangular capillary filled with a bacterial suspension in a magnetizable liquid placed in a uniform horizontal magnetic field. The magnetic torque $\textbf{T}_m$ can be exerted on bacteria by creating rod-shaped voids in the magnetizable medium. (b) Snapshots of representative trajectories (up to 20 seconds, PTV) of the bacteria under four different magnetic field strengths ($\phi = 0.1$, scale bars: $\SI{100}{\micro\meter}$). The white circles and crosses represent the start and end points of the trajectory tracked, respectively. (c) Swimming direction distributions for four different magnetic field strengths (ca. 100 bacteria were tracked for each distribution for ca. 10 seconds each). (d) Nematic order parameter as a function of magnetic field strength for three different ferrofluid concentrations. Error bars indicate the standard deviation of the time-varying order parameter measured over $\SI{10}{\second}$. (e) Bacterial body orientation distributions for four different magnetic fields. Blue lines indicate best fits of $Ae^{-\beta\sin^2(\theta^{\prime})}$. (f) The inverse of body orientation fluctuations as a function of magnetic field strength. Error bars represent the standard deviation of the time-varying square root $\beta$ measured over $\SI{10}{\second}$, where $\beta$ is determined from the fit to the cumulative distribution per second. Dashed lines indicate linear fits.}\label{fig1}
\end{figure*}

\begin{figure*}[t]
\begin{center}
\includegraphics[width=180mm]{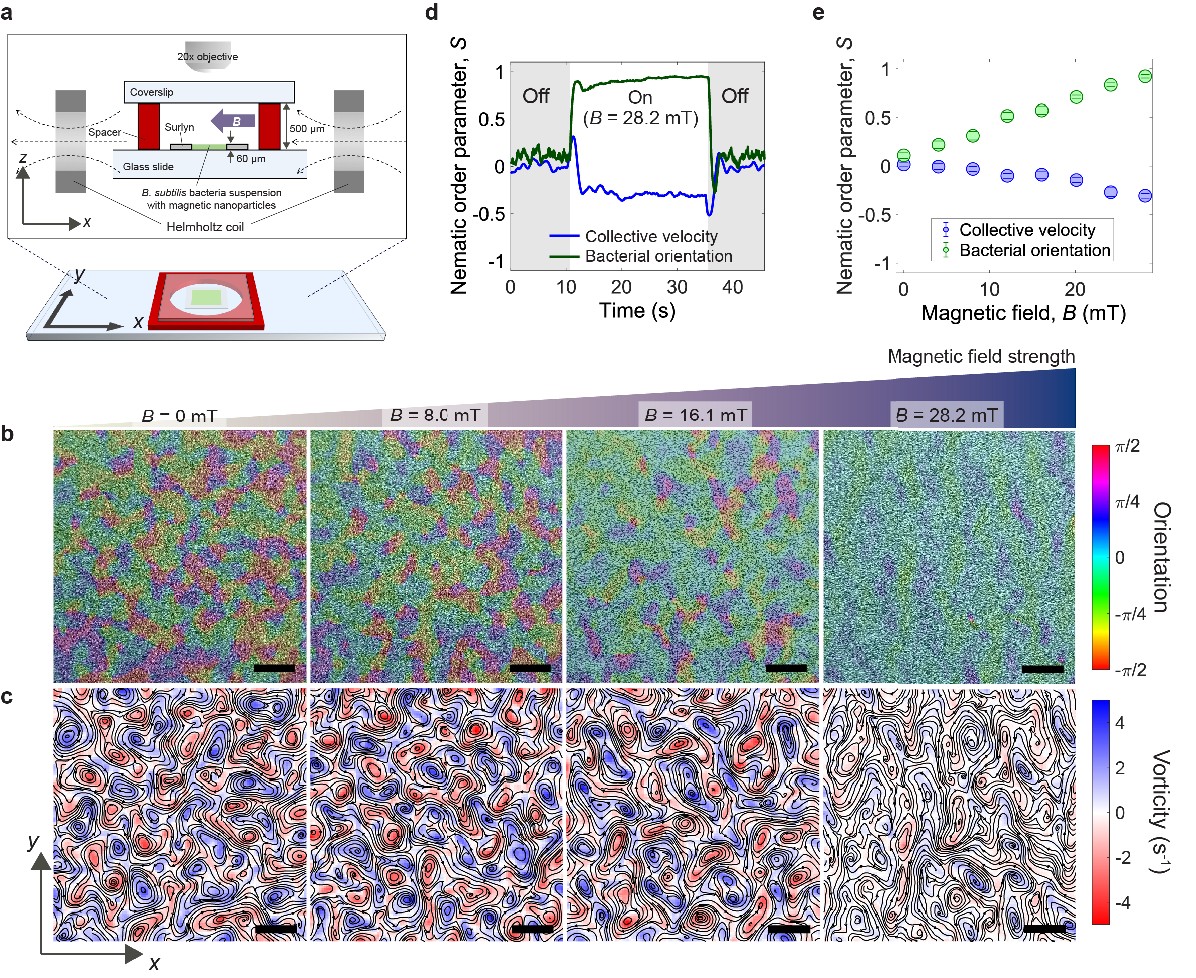}
\end{center}
\caption{\textbf{Magnetic control of orientational order and collective velocity in bacterial turbulence.} (a) A schematic illustration of the experimental setup. (b) Representative snapshots of bright-field images overlapped with the orientation field for four different magnetic fields and (c) the corresponding vorticity fields with the velocity streamlines ($\phi = 0.1$, scale bars: $\SI{100}{\micro\meter}$). The orientation field and velocity field were obtained using the plugin OrientationJ for ImageJ and particle image velocimetry, respectively. (d) Nematic order parameters for bacterial orientation and collective velocity as a function of time during a 25-second long magnetic field pulse. (e) Nematic order parameters as a function of magnetic field strength. }\label{fig2}
\end{figure*}

In this article, we show a method for exerting uniform torques on large populations of non-spherical active agents by immersing them in a magnetic liquid medium and generating the torque with an external magnetic field. We demonstrate this with \textit{Bacillus subtilis} 3610, archetypical bacteria forming active turbulent states. We show that the swimming orientation and nematic ordering can be controlled magnetically not only in dilute but also in dense, turbulent suspensions. We show further that the nematically aligned dense suspension is unstable as the active stresses from the bacteria induce orientational undulation. The length scale of the undulation is independent of the magnetic field strength, providing conclusive experimental evidence of the intrinsic hydrodynamic instability in bacterial turbulence that is further supported by a two-field continuum model. 

\section*{Results}
The magnetically controllable bacterial suspensions were prepared by mixing \textit{B. subtilis} cultured in Terrific Broth (TB) medium with polyethylene glycol (PEG) stabilized ferrofluid (Ferrotec, PBG300). The volumetric concentration $\phi$ of the ferrofluid was varied from 0.02 to 0.10 (i.e., from 2\% to 10\%). The presence of the magnetic nanoparticles, stabilizing agent, or dilution of the TB medium did not affect the mean swimming velocity of the bacteria (Fig. S1). Furthermore, as desired, the nanoparticles were observed not to attach to the bacteria (Fig. S2). The suspensions were studied under a uniform magnetic field generated by a horizontal Helmholtz coil (Fig. \ref{fig1}a).

\subsection*{Magnetic control of the alignment of individual bacteria}
At low bacterial densities ($c_0 \approx 1\sim\SI{2e7}{cells\per\centi\meter^3}$) and in the absence of magnetic field, the bacteria swim at speeds of $\sim \SI{20}{\micro\meter\per\second}$ in an isotropic manner as expected. When a uniform magnetic field is applied, the bacteria align with the field in a nematic manner, and the nematic order increases with increasing magnetic field strength (Fig. \ref{fig1}b and Supplementary Videos 1 and 2), which can be characterized by the velocity orientation distribution of tracked bacteria (Fig. \ref{fig1}c). 
The degree of the nematic alignment can be quantified by using the nematic order parameter
\begin{equation} \label{eq1}
S = \langle \cos2(\theta_j(t)-\theta_B) \rangle_{j,t}
\end{equation}
where $\theta_j(t)$ denotes the orientation of the velocity (${\bf v}_b$) of $j$th bacterium at time $t$, $\theta_B = \pi$ is the direction of the magnetic field, and $\langle \cdot \rangle_{j,t}$ indicates ensemble average over all tracked bacteria and temporal durations of the tracks ($\SI{10}{\second}$) measured after the new steady-state has been reached after turning on the magnetic field. As shown in Fig. \ref{fig1}d, the nematic order increases with the magnetic field strength and approaches unity at $\phi = 0.1$ already under a modest field strength of 10 mT. 

To clarify the mechanism of the magnetic orientation, we analyzed the bacterial body orientations, ${\bf n}_b = (\cos\theta^{\prime},\sin\theta^{\prime})$, by using ellipsoidal fits (Fig. \ref{fig1}e). The cell body orientation distribution is well fitted by $A\exp[-\beta\sin^2(\theta^{\prime})]$ where $A$ is a normalization factor, and $\beta$ is a fitting parameter, suggesting that the bacterial body orientations follow the alignment mechanism described by the nematic potential, $-\sin^2(\theta^{\prime})$ (see Materials and Methods for details). This potential can be derived from the interaction of the uniform magnetic field with the effective magnetic moment of a non-magnetic void in the ferrofluid\cite{Wang}. The fitting parameter corresponds to the inverse of the orientational fluctuation, i.e., $\beta=1/\langle\delta\theta^{\prime2}\rangle$ that we found to increase almost linearly with the magnetic field strength (Fig. \ref{fig1}f)\cite{Wang}. 

At the highest magnetic field strengths and longest durations of observation, the magnetic nanoparticles start to form chains as expected from dipolar forces (Fig. \ref{fig2}b and Supplementary Video 2). However, the chaining occurs much more slowly than the directional change in bacterial orientation. Once the magnetic field is turned 
off, the nanoparticle chains redisperse quickly into a homogeneous isotropic dispersion.

\subsection*{Magnetic control of bacterial turbulence}
At high bacterial densities $c_0 \approx \SI{6e10}{cells\per\centi\meter^3}$ the active turbulence appeared in the magnetic bacteria suspensions similarly as in the regular non-magnetic suspension (Figs. \ref{fig2}a,b). The addition of the magnetic nanoparticles did not significantly change the collective velocity or the intrinsic vortex structure, compared to the control experiments done without magnetic nanoparticles (Figs. S3a,b). In contrast to the dilute samples studied in glass capillaries, the dense suspensions were investigated as thin films $\sim \SI{60}{\micro\meter}$ high with a large liquid-air interface to allow the \textit{B. subtilis} bacteria to access oxygen to maintain their motility (Fig. \ref{fig2}a). When a magnetic field was applied, the disordered turbulent state was maintained at low field strengths, while in stronger magnetic fields the bacteria aligned with the direction of the applied field (Fig. \ref{fig2}b and Supplementary Video 3). This is in contrast to the dilute system (Fig. \ref{fig1}), where the nematic alignment begins to increase even at the lowest magnetic fields (Fig. \ref{fig1}d), suggesting that the active turbulent states can resist some degree of external torque. 

Peculiarly, additional transverse flows appeared in high magnetic fields (Fig. \ref{fig2}c and Supplementary Video 4). The time series of the nematic order parameters for the bacterial orientation and collective velocity are plotted in Fig. \ref{fig2}d. When the magnetic field is off, both order parameters are close to zero. However, at high field strengths, the order parameter for the bacterial orientation sharply increases up to about unity, whereas that for the collective velocity drops down to about $-0.3$. When the field is turned off again, both values approach zero, and in turn, the turbulent state is recovered -- demonstrating the ability to switch the system between nematic order and active turbulence magnetically. The tendencies of nematic ordering and transverse flows were found to increase monotonically with the magnetic field strength (Fig. \ref{fig2}e).

\begin{figure*}[t]
\begin{center}
\includegraphics[width=180mm]{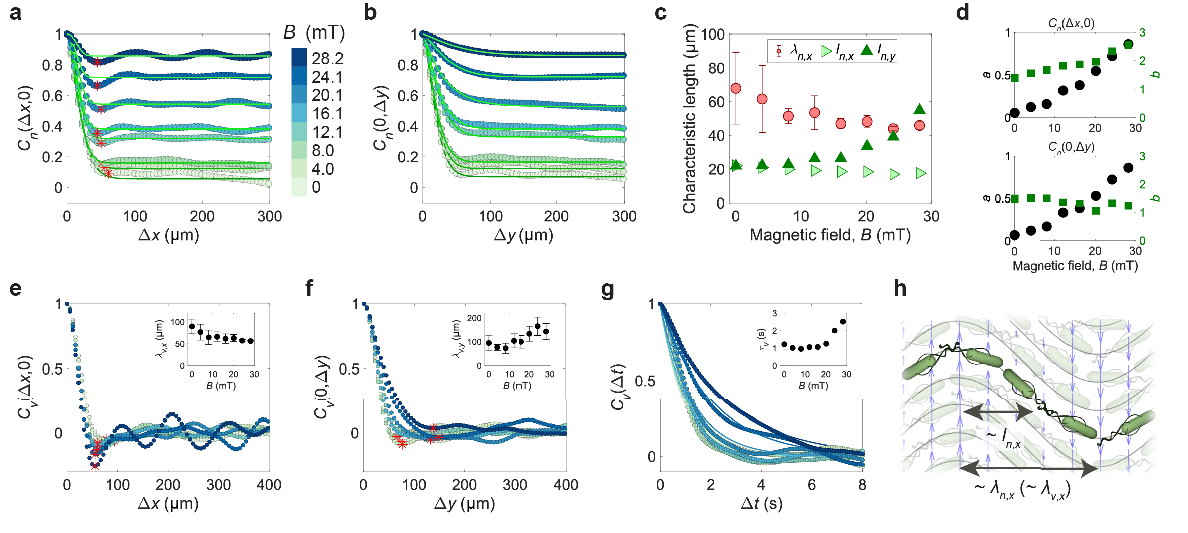}
\end{center} 
\caption{\textbf{Correlations in magnetically tuned bacterial turbulence.} (a,b) Spatial correlogram of bacterial orientation (a) parallel ($x$) and (b) perpendicular ($y$) to the magnetic field. Green lines indicate fits according to equation \eqref{eq3}, and red asterisks indicate the first local minima used to estimate a characteristic length $\lambda_{n,x}$. (c) Characteristic lengths as a function of magnetic field strength. Error bars indicate standard deviations for a time series \SI{10}{\second} long. (d) Fitting parameters, $a$ and $b$ as a function of magnetic field strength. (e) and (f) The normalized velocity correlation in the directions parallel ($x$) and perpendicular ($y$) to the magnetic field. The insets display characteristic lengths obtained from local minima (red asterisks). Error bars represent standard deviations for a duration \SI{10}{\second}. (g) The temporal normalized velocity correlation. The solid lines indicate fitting curves of $e^{-\Delta t/\tau}$, The characteristic time is plotted in the inset. For (a), (b), and (e)-(g), all the fitting is carried out within two-thirds of the system size ($\SI{621.6}{\micro\meter}$) or duration ($\SI{10}{\second}$) to ensure a sufficient number of ensemble averages. (h) Schematic illustrations of the undulation hydrodynamic instability and the defined characteristic length scales.}\label{fig3}
\end{figure*}

In order to quantify the ordered structures in the emergent patterns, we define the correlation function of bacterial orientation as follows:
\begin{equation} \label{eq2}
C_n(\Delta{\bf r}) = \left\langle\langle \cos2(\theta({\bf r}+\Delta{\bf r},t)-\theta({\bf r},t)) \rangle_{{\bf r}}\right\rangle_t
\end{equation} 
where $\Delta{\bf r}$ is a distance, and $\langle \cdot \rangle_{{\bf r},t}$ denotes an ensemble average in space and time (averaged over 10 seconds). To clarify the anisotropic structural details in the orientational ordering, we decompose the correlation into $\Delta x$ and $\Delta y$ components (Figs. \ref{fig3}a,b). Depending on the magnetic field, these correlations display extension of correlated area in both parallel ($x$) and perpendicular ($y$) directions. The orientational correlation can be fitted by
\begin{equation} \label{eq3}
G(|\Delta{\bf r}|) = (1 - \textit{a})\textit{e}^{-\left(\frac{|\Delta{\bf r}|}{\textit{l}}\right)^\textit{b}} + \textit{a}
\end{equation} 
where $l$, $a$, and $b$ are adjustable fitting parameters \cite{Cvetko}. The parameter $l$ denotes the coherence length that corresponds to the average nematic domain length and is plotted for each of the directions parallel and perpendicular to the applied field as a function of the magnetic field strength in Fig. \ref{fig3}c. While the coherence length in the $x$ direction is kept almost constant over the magnetic field, that in the $y$ direction shows a gradual increment. 

Importantly, a slight yet wavy profile appears in the $x$ direction. To extract the wavelength of the bending, we define a characteristic length $\lambda$ as the first local minimum value (Figs. \ref{fig3}c,h). The length $\lambda_{n,x}$ is nearly constant as is the coherence length $l_{n,x}$. The quantity $a$ stands for the degree of order, and $b$ for the distribution width of $l$ (Fig. \ref{fig3}d). The nematic order monotonically increases in both $x$ and $y$ directions (Fig. \ref{fig3}d top and bottom), consistent with Fig. \ref{fig2}d. On the other hand, the width of the distribution of $l$ increases in the $x$ direction but decreases in the $y$ direction, indicating that the horizontal nematic domain becomes more prominent along with the magnetic field strength. Together, the above observations suggest that the nematic order over the entire region is enhanced by the magnetic field, but there exists a wavelike orientational structure with a constant length scale along the magnetic field direction in the nematically ordered state. 

To quantify the characteristics of the flow fields, we analyzed the normalized velocity correlation in space defined as
\begin{equation} \label{eq4}
C_v(\Delta{\bf r}) = \left\langle\frac{\langle{\bf v}({\bf r}+\Delta{\bf r},t)\cdot{\bf v}({\bf r},t) \rangle_{{\bf r}}}{\langle {\bf v}({\bf r},t)\cdot{\bf v}({\bf r},t) \rangle_{{\bf r}}}\right\rangle_t
\end{equation} 
where the ensemble average is carried out as above. Figures \ref{fig3}e and f show the velocity correlation in the $x$ and $y$ directions, respectively. Since the flow field includes vortical structures with clockwise and anti-clockwise handedness, the correlation function takes a local minimum, and hence we can obtain characteristic lengths $\lambda_{v,x}$ and $\lambda_{v,y}$ in both directions, as with $\lambda_{n,x}$. The insets in Figs. \ref{fig3}e and f indicate the $B$-dependence of the characteristic lengths, which is consistent with those of bacterial orientation. Notably, the length $\lambda_{v,x}$ is comparable to $\lambda_{n,x}$, suggesting that the periodic longitudinal flow is attributed to the periodic undulating orientation field of bacteria.

Furthermore, we analyzed the temporal normalized correlation function of velocity defined as follows: $C_v(\Delta t) = \langle{\bf \hat{v}}({\bf r},t+\Delta t)\cdot{\bf \hat{v}}({\bf r},t) \rangle_{{\bf r},t}$, where ${\bf \hat{v}}$ is a unit vector of the velocity field, and the ensemble average is taken as above. By fitting it with an exponential function, $\exp[-\Delta t/\tau]$, we can obtain a typical correlation time (Fig. \ref{fig3}g). In a weak or intermediate magnetic field, the time $\tau_v$ is approximately $\SI{1}{\second}$ and corresponds to a typical lifetime of turbulent vortices\cite{Dombrowski}, but in the strong magnetic field, it increases by a factor of two due to the long persistence of the nematic ordered phase.

\begin{figure*}[t]
\begin{center}
\includegraphics[width=180mm]{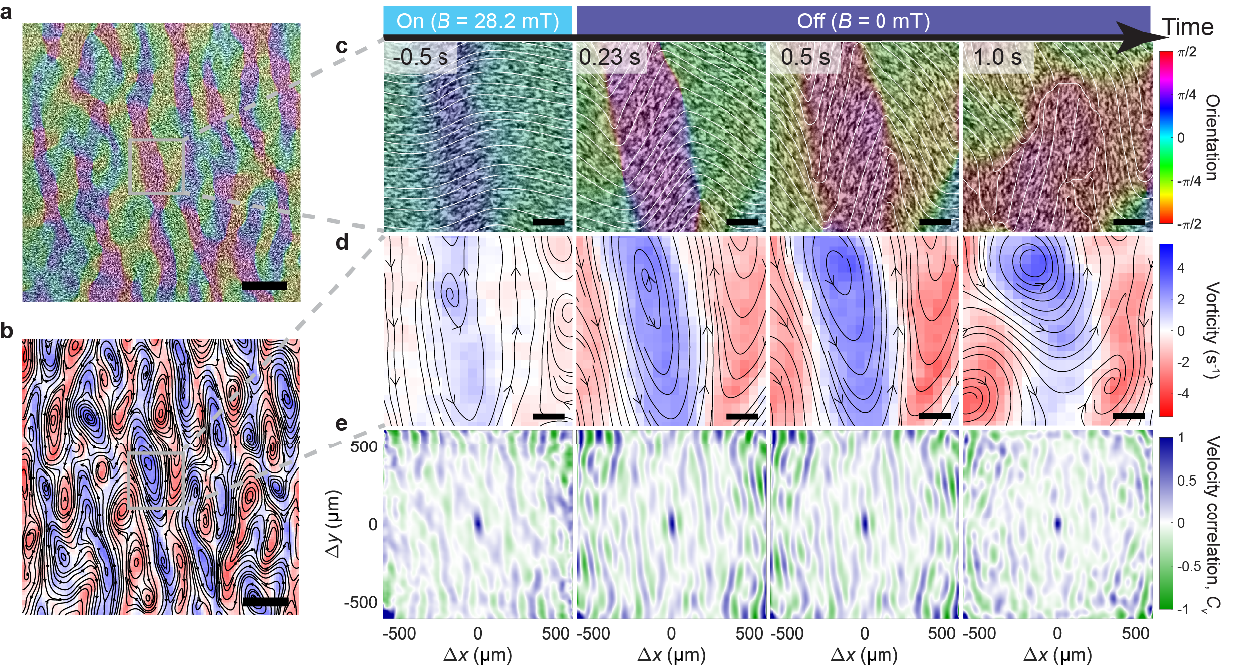}
\end{center}
\caption{\textbf{Instability of the nematically aligned state and recovery of the active turbulence.} (a) and (b) Representative snapshots of the bright-field image overlaid with bacterial orientation map (a) and vorticity field with velocity streamlines (b), acquired at $\SI{0.5}{\second}$ after the magnetic field $B =28.2$ mT is turned off. Scale bars, $\SI{100}{\micro\meter}$. (c) and (d) Time series of snapshots of the bacterial orientation with orientation contours and vorticity field inside the gray rectangle in (a) and (b), respectively. Scale bars, $\SI{20}{\micro\meter}$. (e) Time series of spatial normalized velocity correlation. At $\SI{0.5}{\second}$, the characteristic length of undulation is defined as $\lambda_{n,x} \sim \SI{55.3}{\micro\meter}$ from the first local minimum. }\label{fig4}
\end{figure*}

\subsection*{Turning the active turbulence on and off magnetically: intrinsic hydrodynamic instability}
Since the external magnetic field and torque can be switched on and off quickly, our magnetic approach allows probing transitions between different collective states. Such control is impossible using, e.g., geometric boundaries that are stationary\cite{Wioland1,Wioland2,Wioland3,Beppu1,dogic,Nishiguchi2,hardouin,Beppu2}. In the case of the nematically aligned dense \textit{B. subtilis} population, turning off the magnetic field leads to the rapid growth of the underlying minute undulation (Figs. \ref{fig4}a-d and Supplementary Video 5). 
The analysis of the spatial velocity correlations as a function of time shows that the stripe-like correlation becomes more prominent at short time scales ($< \SI{0.5}{\second}$) after the magnetic field is turned off (Fig. \ref{fig4}e). At longer time scales ($> \SI{1}{\second}$), the full-fledged active turbulence is recovered.

\begin{figure}[t]
\begin{center}
\includegraphics[width=88mm]{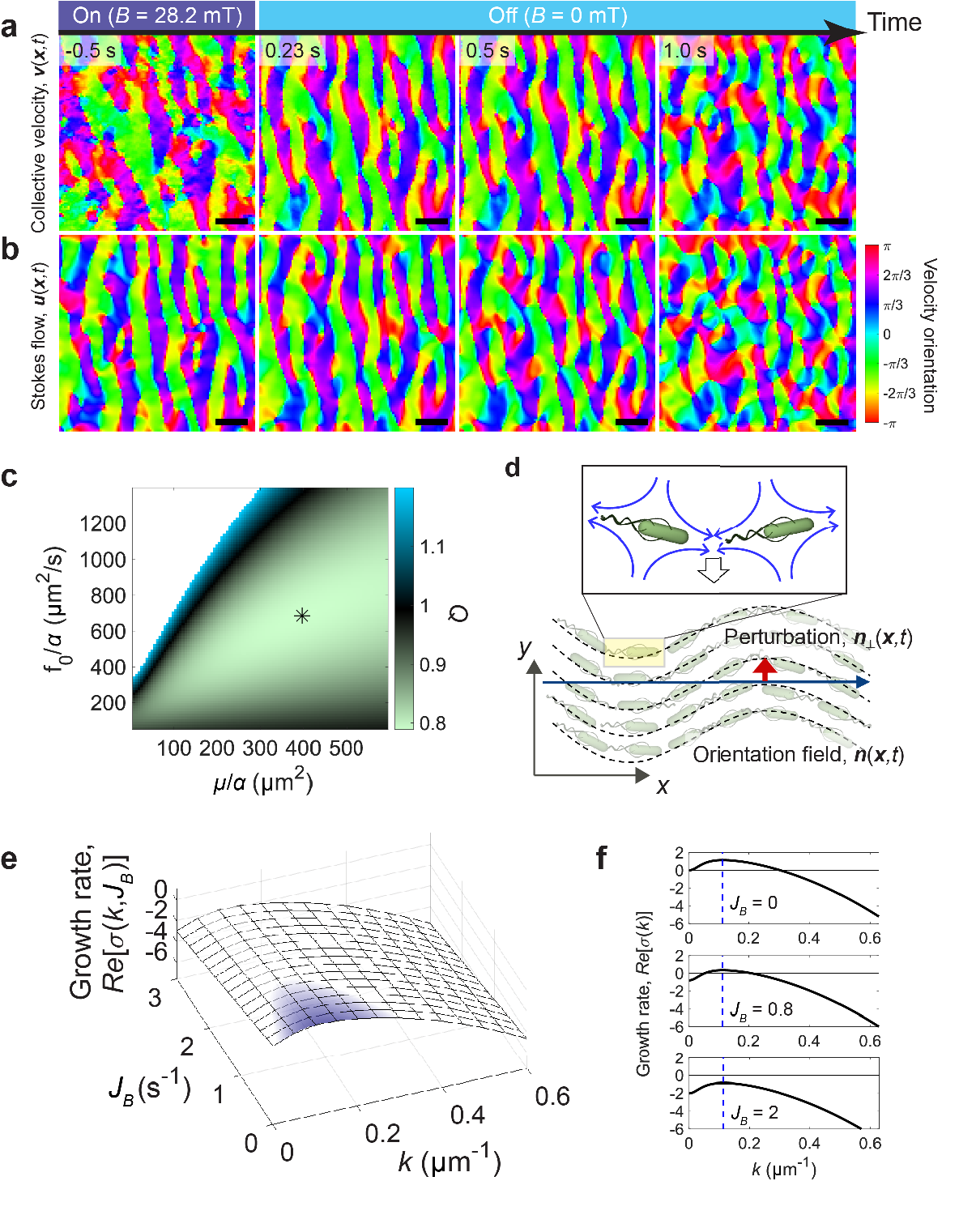}
\end{center}
\caption{\textbf{Intrinsic hydrodynamic instability in bacterial turbulence.} (a) and (b) The time series of velocity orientations of collective velocity and fluid flow velocity right before and after the magnetic field ($B = 28.2$ mT) is turned off. The fluid flow velocities are calculated from the Stokes equation \eqref{eq6} where experimental data are used for bacterial orientations. Scale bars, $\SI{100}{\micro\meter}$. (c) The quality function $Q$ calculated from a difference between ${\bf u}({\bf r},t)$ and ${\bf v}({\bf r},t)$. The black asterisk denotes a minimum point of $Q$. (d) Schematic illustrations of the undulation hydrodynamic instability and the definition of orientation perturbation in a nearly aligned state. (e) The real part of the growth rate \eqref{eq11} as a function of wavenumber $k$ and magnetic field coefficient $J_B$. The used parameters are $\gamma = 0.9$, $f_0/\alpha = \SI{686}{\micro\meter^2\per\second}$, $\mu/\alpha = \SI{397}{\micro\meter^2}$, and $D_n = \SI{17.3}{\micro\meter^2\per\second}$. The positive growth rate is colored blue. (e) The growth rate against wavenumber $k$ at three different values of $J_B$. The transverse broken blue lines indicate the typical wavenumber at which the growth rate is maximized.}\label{fig5}
\end{figure}

As shown in Refs.\cite{Ramaswamy2,Shelley,Zhou,Genkin,Turiv}, swimming bacteria like \textit{B. subtilis} are pusher-type microswimmers that exert force dipoles on their surrounding fluid\cite{Drescher}, and thus the active stress is known to induce orientational undulation, mediated by fluid flows. Such a self-amplifying bending deformation is an important feature of extensile active nematic systems\cite{Chandrakar}. Taking our cue from the activity-induced hydrodynamic instability demonstrated in those earlier works, we investigate whether the transverse flows we observed stem from the active stress by pushers by considering the 2D Stokes equation governing the fluid flow velocity ${\bf u}({\bf r},t)$\cite{Wioland1,Li}:
\begin{equation} \label{eq6}
-\mu\nabla^2{\bf u} + \nabla p + \alpha{\bf u} = -f_0\nabla \cdot {\bf nn}
\end{equation} 
where $\mu$ is the viscosity coefficient, the pressure $p$ is the Lagrange multiplier for the incompressibility condition ($\nabla \cdot {\bf u} = 0$), $\alpha$ term is the effective friction with the substrate, and the right-hand side of the equation represents the active force with the coefficient $f_0$ ($f_0>0$ for pushers) determined by the bacterial orientation field ${\bf n}({\bf r},t)$. We assume the uniformity of density distribution of bacteria over space even when a magnetic field is applied due to the sufficiently high concentration. Accordingly, the strength of active force $f_0 = qc_0$, where $q$ and $c_0$ are the strength of dipoles and the number density of bacteria, respectively, is assumed to be constant. As shown in Figs. \ref{fig5}a,b, we calculated the fluid velocity field ${\bf u}({\bf r},t)$ from the instantaneous orientation field ${\bf n}({\bf r},t)$ by following equation \eqref{eq6}, which is similar to the collective velocity from PIV analysis ${\bf v}({\bf r},t)$. This suggests that the emergent transverse flow originates from active stress. The optimal parameters in equation \eqref{eq6} are searched in a manner proposed in ref.\cite{Li}. We introduce a parameter $Q$ defined as $Q(\mu/\alpha, f_0/\alpha) = \langle|{\bf u}({\bf r},t) - {\bf v}({\bf r},t)|^2/| {\bf v}({\bf r},t)|^2 \rangle_{{\bf r},t}$, where $\langle\cdot\rangle_{{\bf r},t}$ denotes the spatial and time averages, in order to quantify the extent to which the calculated flow field ${\bf u}({\bf r},t)$ coincides with the experimentally obtained ${\bf v}({\bf r},t)$. The minimum value of $Q$, averaged for $\SI{0.5}{\second}$ after the magnetic field ($B = 28.2$ mT) is off, yields the optimal parameter set $(\mu/\alpha, f_0/\alpha) = (\SI{397}{\micro\meter^2}, \SI{686}{\micro\meter^2\per\second})$ (Fig. \ref{fig5}c).

Bacterial turbulence is, in general, known to be well reproduced by single-field models, which are justified as long as the active forces are sufficiently small compared to viscous ones\cite{Alert,Wensink,Reinken}. In order to quantitatively illustrate the undulation mechanism of nematic-aligned state by fluid flows that active stress induces (Fig. \ref{fig3}h and Fig. \ref{fig5}d), however, we here describe the system using the two-field continuum model with reference to earlier works (Refs.\cite{Wioland1,Wolgemuth,Ramaswamy2,Shelley,Reinken,Reinken2}). The evolution equation of the bacterial orientation field ${\bf n}({\bf r},t)$ can be given in the following:

\begin{equation} \label{eq7}
\begin{split}
\frac{\partial {\bf n}}{\partial t} + ({\bf u} + V_0{\bf n})\cdot\nabla{\bf n} &= D_n\nabla^2{\bf n} + ({\bf I} - {\bf nn})\cdot(\gamma{\bf E}+ {\bf W})\cdot{\bf n}\\
& + J_B({\bf h}\cdot{\bf n}){\bf h}\cdot({\bf I}-{\bf nn}).
\end{split}
\end{equation} 
On the left-hand side, the bacterial alignment is advected by ${\bf u} + V_0{\bf n}$ where $V_0$ is the swimming speed of bacteria. On the right-hand side, the first term denotes the diffusion with the coefficient $D_n$ through nematic interactions due to the rod-like shape of bacteria. The second term reorients the bacterial alignment via fluid flow velocity by solvent strain tensor ${\bf E} = (\nabla^T{\bf u} + \nabla{\bf u}^T)/2$ and vorticity one ${\bf W} = (\nabla^T{\bf u} - \nabla{\bf u}^T)/2$, with a shape parameter $-1\leq\gamma\leq1$. As discussed in the case of dilute suspensions (Fig. \ref{fig1}), bacteria are aligned with respect to the magnetic field, and hence we incorporate the nematic magnetic torque with the strength $J_B$ and the unit vector of the field ${\bf h}$ in the last term.

To understand the undulation instability and its magnetic controllability, we here analyzed the linear stability of a nearly aligned state along the magnetic field direction. The presence of the magnetic field that forces bacteria to align in parallel to the $x-y$ plane allows us to simply consider a two-dimensional system, where the magnetic field direction is set to ${\bf h}=(1, 0)$, and only perturbations in the $y$ direction: ${\bf n}({\bf r},t) = {\bf e}_x + {\bf n_{\perp}}$ ($|{\bf n_{\perp}}|\ll1$), where ${\bf e}_x=(1, 0)$ and ${\bf e}_x\cdot{\bf n_{\perp}} = 0$ (Fig. \ref{fig5}d). Fourier transformation of the orientation field, ${\bf n_{\perp}}({\bf r}) = {\bf \tilde{n}}({\bf k})e^{i{\bf k}\cdot{\bf r} + \sigma t}$, yields the real part of the growth rate along the $x$ direction (see Materials and Methods for details):
\begin{equation} \label{eq11}
\mathrm{Re}[\sigma] = \frac{f_0k^2}{2(\alpha + \mu k^2)}(\gamma + 1) - D_nk^2 - J_B.
\end{equation} 
As for the characteristic lengths of undulation, we can obtain a typical wavenumber that maximizes the obtained growth rate \eqref{eq11} by differentiating it with respect to $k$:
\begin{equation} \label{eq12}
k_c = \left(\frac{-D_n + \sqrt{D_nF}}{D_nM}\right)^{\frac{1}{2}},
\end{equation}
where $F = f_0(1+\gamma)/(2\alpha)$, $M = \mu/\alpha$, and we set $\gamma$ to 0.9 due to the rod shape of bacteria\cite{Wioland1}. In the experiment, we extract the characteristic length $\lambda_{n,x}$ from the local minimum point in the velocity correlation at $\SI{0.5}{\second}$ after the magnetic field is off, which is $\lambda_{n,x} = \SI{55.3}{\micro\meter}$. This corresponds to $2\pi/k_c$, thus giving the estimation $D_n = \SI{17.3}{\micro\meter^2\per\second}$. Using the parameters estimated above, the growth rate is mapped against $k$ and $J_B$ in Figs. \ref{fig5}e,f. At a small $J_B$, the presence of a positive regime of $\sigma$ in a long wavelength suggests that a nematic-aligned state becomes unstable, resulting in undulation instability. In contrast, at a large $J_B$, the unstable mode of orientation perturbations disappears, as the growth rate is suppressed by $J_B$. The transition point above which a nematic-aligned state is stable is given by
\begin{equation} \label{eq13}
J_B \geq \frac{1}{M}\left(\sqrt{F} - \sqrt{D_n}\right)^2.
\end{equation}
$J_B$ indicates the strength of nematic alignment by the magnetic field. Based on the microscopic description (see Materials and Methods for details), $J_B$ can be given by $J_B = 2\Gamma\epsilon_1 B^2$ where $\Gamma$ represents the response of bacterial orientation to magnetic fields and $\epsilon_1$ is a magnetic constant. Given that $\sqrt{2\Gamma\epsilon_1/D_{\theta}}$ (mT$^{-1}$) where $D_{\theta}$ is the rotational diffusion coefficient corresponds to the slope of the inverse orientation fluctuation (0.347 mT$^{-1}$) at a 10\% ferrofluid concentration in Fig. \ref{fig1}f, one can estimate a value of $2\Gamma\epsilon_1$ using $D_{\theta} = \SI{0.066}{\radian^2\per\second}$ obtained from an independent experiment (Fig. S3). Taken together, the transition point $B_c$ can be estimated to be $B_c \approx 12$ mT, which is approximately a point at which the nematic order parameter measured in the experiment exceeds 0.5 (Fig. \ref{fig2}e and Fig. \ref{fig3}d). Importantly, our model assumes the instability from a nearly nematic-aligned state, which is not exactly the behavior observed experimentally, i.e., the transition from bacterial turbulence to the nematic state. This possible discrepancy motivated us to further confirm the transition behavior from a nematic state to bacterial turbulence by implementing the temporal change in the magnetic field (Fig. S5). Programming the linear decrease in the magnetic field strength from $B$ = 28.2 mT to 0 allowed for a gradual transition from an aligned state to a turbulent state. We found that the transition point appears to be still at $B = 10 \sim 20$ mT, as in the case of Fig. \ref{fig2}e.

\section*{Discussion and concluding remarks}
In summary, we have demonstrated the ability to control the swimming direction and collective states of non-magnetic bacteria by exerting externally controllable torques on them through a biocompatible and magnetizable liquid medium. In dilute bacterial suspensions, the torque can constrain the alignment of the rod-shaped bacteria and their swimming direction to the direction of the magnetic field, leading to externally tunable nematic ordering. In dense suspensions, the magnetic torque sculpts the isotropic active turbulent state to a nematically ordered state – however, flows perpendicular to the magnetic field also appear. The nematically aligned configuration is undulated with a characteristic length scale that is almost independent of the applied magnetic field. We put forward a simple continuum model and show that linear stability analysis leads to a characteristic length of the undulation that is independent of the magnetic field, in good agreement with the experimental observation. 

Our results suggest that externally controllable torque generation on non-magnetic bacteria via a magnetizable medium is a powerful tool to control both individual bacteria as well as their collective states. In contrast to other techniques such as optical\cite{Rasmussen} or acoustic tweezers\cite{Takatori}, our approach leads to the generation of uniform torques everywhere in large samples of up to cm-scales. We foresee that this will be especially important as large-scale patterns often appear in active systems. In contrast to studies with magnetotactic bacteria\cite{clement}, in our approach, the magnetic nanoparticles are outside non-magnetic bacteria. This will likely allow generalization to all bacterial species and even other micro-organisms with different approaches to motilities and collective states. Finally, dynamic and non-uniform magnetic fields are foreseen to lead to even more advanced spatiotemporal control of both individual bacteria and their collective states via additional translational forces\cite{Jaakko} and programmable time-dependent magnetic fields and torques\cite{Jaakko3}. 


\clearpage

\section*{Materials and Methods}

\subsection*{Preparation of bacterial suspensions in magnetizable media}
\textit{Bacillus subtilis} strain 3610 (Bacillus Genetic Stock Center, Original code: NCIB3610, BGSCID: 3A1) was grown by inoculating a single bacterial colony into \SI{5}{\milli\liter} of Lysogeny broth (LB) medium (created by dissolving a \SI{25}{\gram} LB broth base powder (Invitrogen, 12795027) in \SI{1}{\liter} MilliQ water and autoclaving) in a \SI{50}{\milli\liter} Falcon tube (Falcon, 352070) and incubating it overnight on an orbital shaker (Grant-bio, PSU-10i) inside an incubator (Memmert, IN55) at \SI{37}{\celsius}. Next day, \SI{100}{\micro\liter} of the culture was added to \SI{20}{\milli\liter} of fresh Terrific Broth (TB) medium (Gibco, 11632139) and further incubated on the orbital shaker in the incubator at \SI{37}{\celsius} at 160 RPM for ca. 3 hours. After the optical density of the culture, measured at 600 nm in a plastic cuvette with 10 mm path length using a VIS spectrometer (Thermo Scientific, GENESYS 30), reached ca. $0.4$, the culture was diluted ca. 10 to 40 times in TB medium to create a dilute bacterial suspension. The dense bacterial suspensions were created by centrifuging (Fisher Scientific, GT R1 Centrifuge) the bacterial suspension (obtained after 3 hours of incubation) at 3000 RPM at room temperature for 5 minutes to increase the density of the suspension to $\sim 20\% v/v$ at which bacterial turbulence can be observed. The bacterial concentration was estimated from the optical density measured at \SI{600}{\nano\meter} using the relationship that OD = 1.0 at \SI{600}{\nano\meter} corresponds to $c_0 \approx \SI{7e8}{cells\per\centi\meter^3}$\cite{Janosi}. Finally, the dilute and concentrated bacterial suspensions were mixed with a biocompatible ferrofluid (Ferrotec, PBG300) at a desired proportion: 2\%, 5\%, or 10\%. According to the specifications and physical properties of PBG300 provided by Ferrotec, the viscosity at \SI{27}{\celsius} is \SI{3}{\milli\pascal\second}, and hence we expect the viscosity of the suspension with ferrofluid to be ca. \SI{1}{\milli\pascal\second} even at a 10\% ferrofluid concentration in the absence of magnetic fields, which is the almost same as that of the TB medium.

\subsection*{Sample preparation for microscopy}
The dilute bacterial samples were imaged inside rectangular, 50 mm long glass capillaries 4 mm wide and 0.2 mm deep (with wall thickness 0.2 mm, CM scientific, 3524-050). For the dense bacterial samples, a well was constructed on a regular microscopy glass slide using a thermoplastic ionomer film (DuPont Surlyn, \SI{60}{\micro\meter}) that was attached to the glass slide by heating to \SI{130}{\celsius} on a hotplate. After filling the well, a \SI{500}{\micro\meter} thick spacer with one \SI{2}{\milli\meter} diameter well (Invitrogen P18174, 10276972) was placed around the well and covered from the top with a glass coverslip to prevent the suspension from evaporating. Since the \textit{B. subtilis} bacteria require lots of oxygen, the bacterial motility is highly maintained only within a few tens of microns from the water-air interface, thus allowing us to consider the system to be quasi-two-dimensional.

\subsection*{Microscopy under a uniform magnetic field}
Imaging of the samples under a uniform horizontal magnetic field was done using a setup described earlier with small modifications\cite{Jaakko2}. Briefly, the uniform horizontal field was created using an electromagnet coil pair (GMW 11801523 and 11801524) with a 50 mm gap. The coils were driven with a DC power supply (BK Precision 9205). Microscopy imaging of the sample was done using a custom-made microscope setup consisting of a $20 \times$ objective lens with a numerical aperture of 0.45 (Nikon, TU Plan Fluor), an Epi-Illuminator Module (Thorlabs, CSE2200), and a $0.5 \times$ Camera Tube (Thorlabs, WFA4102) equipped with a CMOS camera (XIMEA, MC050MG-SY-UB). The sample was illuminated from below in transmitted light configuration using a light-emitting diode light source (Thorlabs, MNWHL4) with a collimator which is diffused right before it hits the sample. The images were acquired at 30 frames per second with an exposure time of \SI{33}{\milli\second}.

\subsection*{Data processing and analysis}

The trajectories of individual swimming bacteria were tracked with Particle Tracking Velocimetry (PTV) analysis by using the TrackMate plugin \cite{Ershov} in Fiji\cite{Schindelin}. Prior to PTV analysis, raw images from the experiments were post-processed by subtracting the background, enhancing the contrast, smoothing by a Median filter, and inverting the image intensity histogram. Immotile bacteria, such as those adsorbed on the glass capillary surface, were eliminated from the trajectories tracked for over \SI{2}{\second} by accepting only trajectories that showed ballistic motion with mean square displacement in the regime of \SI{1}{\second} proportional to $t^c$ with $c$ greater than 1.5. In addition, to reduce the noise in the velocity determination, the accepted trajectories were smoothed by taking the difference between two data points along the trajectory separated by 10 frames (corresponding to the typical time for a bacterium to move one cell's body length), instead of considering neighboring data points. These analyses, with the exception of the tracking, were conducted in Matlab (MathWorks) using custom-made scripts.

The velocity field of the bacterial collective motion was obtained using Particle Image Velocimetry (PIV) analysis done using the PIVLab toolbox\cite{Thielicke} in Matlab. The Wiener2 denoise filter was used in the image pre-processing, and the interrogation window size was chosen to be $16 \times 16$ pixels$^2$ ($5.52 \times \SI{5.52}{\micro\meter^2}$), corresponding to the typical body length of the studied bacteria. To reduce the noise, the acquired velocity field was further smoothed by averaging over $\SI{1}{\second}$, which corresponds to the typical lifetime of turbulent vortices (Fig. \ref{fig3}g).

The local orientation of the bacteria was obtained by using OrientationJ plugin\cite{Rezakhaniha} in ImageJ that detects the direction of the largest eigenvector of the structure tensor of the image. We set the local window size of the structure tensor to $15 \times 15$ pixels (corresponding to $\SI{5}{\micro\meter}$). The orientation field was further coarse-grained and reduced to the same matrix size as that of the PIV velocity field after the convolution with half of the PIV window size. 

The flow velocity field $\textbf{u}$ in equation \eqref{eq6} was computed by a spectral method that first solves in Fourier space and then transforms to real space.

\section*{Theoretical model}
\subsection*{Orientational dynamics of bacteria in a ferrofluid}
Bacteria suspended in a ferrofluid under the application of magnetic fields create rod-shaped voids in which magnetic moments anti-parallel to the field are induced\cite{Wang}. Their interactions nematically reorient bacteria along the magnetic field. The orientational dynamics of bacteria in a dilute suspension can be described by
\begin{equation} \label{eq14}
\frac{d \theta}{dt} = -\Gamma\frac{\partial U_m}{\partial \theta} + \sqrt{2D_{\theta}}\eta(t).
\end{equation} 
The first term is the magnetic torque with the response coefficient $\Gamma$ of bacterial orientation to the magnetic field and the dimensionless magnetic potential $U_m = \epsilon_1 B^2\sin^2\theta + \epsilon_2 B^2$ where $\epsilon_{1,2}$ are magnetic constants dependent on the magnetic susceptibility, the permeability of vacuum, and the volume of the bacterium\cite{Wang}. The second term is the white Gaussian noise with the rotational diffusion coefficient that satisfies $\langle\eta(t)\rangle=0$ and $\langle\eta(t)\eta(t^{\prime})\rangle=\delta(t-t^{\prime})$ where $\delta(t-t^{\prime})$ is a Dirac-delta function. Note that tumbling occurs at most once during the observation time of \SI{10}{\second}, and its effect is considered small enough on the time scale of the observation time to be ignored here for simplicity. According to the conventional procedure upon the assumption that the distribution of bacterial orientation is spatially uniform\cite{Beppu1,Peruani}, one can obtain the Fokker-Planck equation of the probability distribution of bacteria heading $\theta$:
\begin{equation} \label{eq15}
\frac{\partial P}{\partial t} = D_{\theta} \frac{\partial^2 P}{\partial \theta^2} + \frac{\partial}{\partial \theta} \left(\Gamma\frac{\partial U_m}{\partial \theta} P\right).
\end{equation} 
The solution in the steady state can be written as
\begin{equation} \label{eq16}
P(\theta) = A\exp\left(-\frac{\Gamma}{D_{\theta}}\epsilon_1 B^2\sin^2\theta\right)
\end{equation} 
where $A$ denotes a normalization factor. This function is fitted to the probability distribution in Fig. \ref{fig1}e. Since this is interpreted as the Gaussian function around $\theta=0$, as shown in Fig. \ref{fig1}f, we analyzed the orientation fluctuation by comparing $e^{-\theta^2/(2\langle\delta\theta^2\rangle)}$ with equation \eqref{eq16}. Thus, $1/\langle\delta\theta^2\rangle$ corresponds to $2\Gamma \epsilon_1 B^2/D_{\theta}$.

To estimate the rotational diffusion coefficient $D_{\theta}$, we tracked swimming trajectories in a dilute bacterial suspension with a 10\% ferrofluid concentration using PTV analysis. Here, due to the difficulty in tracking bacterial polarity for a long time, let us use the velocity unit vector instead of bacterial orientation, which could be justified by the fact that polarity and velocity are almost identical in the absence of interactions with others. We analyzed mean square angular displacement, defined by $\langle(\textit{\textbf{d}}(\Delta{\it t})-\textit{\textbf{d}}(0))^2\rangle$ where $\textit{\textbf{d}}(\Delta{\it t})$ denotes a unit vector of swimming velocity $\Delta{\it t}$ after an initial condition $\textit{\textbf{d}}(0)$ (Fig. S4). The ensemble average $\langle\cdot\rangle$ was taken for all the bacteria which were tracked for over $\SI{10}{\second}$ and exhibited straight trajectories with mean curvature radii of more than $\SI{700}{\micro\meter}$ to rule out bacteria showing apparent curved trajectories due to their chirality and hydrodynamic interactions with interfaces. Fitting it by $2(1-\exp(-D_{\theta}\Delta{\it t}))$ yields the estimate of $D_{\theta} = \SI{0.066}{\radian^2\per\second}$\cite{Beppu2}.

\subsection*{Linear stability analysis}
To reveal the undulation instability, we analyzed the linear stability of a uniformly aligned state along the magnetic field factor ${\bf J_B} = J_B{\bf h}$ with the coefficient $J_B$ related to the magnetic field strength and the direction ${\bf h} = (1, 0)$. We consider perturbations for the orientation field in the $y$ direction: ${\bf n}({\bf r},t) = {\bf e}_x + {\bf n_{\perp}}$ ($|{\bf n_{\perp}}|\ll1$) where ${\bf e}_x=(1, 0)$ and ${\bf e}_x\cdot{\bf n_{\perp}} = 0$. We also define perturbations for the fluid flow velocity and pressure, ${\bf u}({\bf r}) = {\bf u^{\prime}}$ ($|{\bf u^{\prime}}|\ll1$), and $p({\bf r}) = \eta$ ($\eta\ll1$), respectively. Substituting these variables for equations \eqref{eq6} and \eqref{eq7}, and retaining first-order terms of the infinitesimal quantities, equations \eqref{eq6} and \eqref{eq7} are readily reduced to the following equations:
\begin{equation} \label{eq8s}
-\mu\nabla^2{\bf u^{\prime}} + \nabla \eta + \alpha{\bf u^{\prime}} = -f_0\nabla \cdot ({\bf n_{\perp}}{\bf e}_x + {\bf e}_x{\bf n_{\perp}}),
\end{equation} 

\begin{equation} \label{eq9s}
\begin{split}
\frac{\partial {\bf n_{\perp}}}{\partial t} + V_0{\bf e}_x\cdot\nabla{\bf n_{\perp}} &= D_n\nabla^2{\bf n_{\perp}} + ({\bf I} - {\bf e}_x{\bf e}_x)\cdot(\gamma{\bf E^{\prime}}+ {\bf W^{\prime}})\cdot{\bf e}_x\\
& - J_B{\bf n_{\perp}}.
\end{split}
\end{equation} 
Upon the incompressibility condition ($\nabla \cdot {\bf u^{\prime}} = 0$), Fourier transformation of the fluid flow velocity, ${\bf u^{\prime}}({\bf r}) = {\bf \tilde{u}}({\bf k})e^{i{\bf k}\cdot{\bf r}}$, gives the solution of the fluid flow velocity in Fourier space:
\begin{equation} \label{eq10s}
{\bf \tilde{u}}({\bf k}) = -\frac{if_0}{\alpha + \mu k^2} \left({\bf I} - \frac{{\bf k}{\bf k}}{k^2}\right)\cdot({\bf n_{\perp}}{\bf e}_x + {\bf e}_x{\bf n_{\perp}})
\end{equation} 
where $k = |{\bf k}|$. We assume the direction of ${\bf k}$ to be an elevation angle $\varphi$, i.e., ${\bf k} = (k\cos\varphi, k\sin\varphi)$. Substitution of the velocity \eqref{eq10s} for equation \eqref{eq9s} and Fourier transformation of the orientation field, ${\bf n_{\perp}}({\bf r}) = {\bf \tilde{n}}({\bf k})e^{i{\bf k}\cdot{\bf r} + \sigma t}$, gives the growth rate:
\begin{equation} \label{eq11s}
\sigma = \frac{f_0k^2}{2(\alpha + \mu k^2)}\cos2\varphi(\gamma\cos2\varphi + 1) - D_nk^2 - J_B - ikV_0\cos\varphi.
\end{equation} 
The real part of the growth rate in the $x$ direction ($\varphi = 0$) reads equation \eqref{eq11}. In addition, the real part of the growth rate in the $y$ direction ($\varphi = \pi/2$) is negative throughout the wavenumber. This means that as long as the system is nearly aligned state in the direction parallel to the magnetic field, the alignment is stable in the transverse direction, which is consistent with the increment in the characteristic lengths in the transverse direction beyond the transition point of the magnetic field (Fig. \ref{fig3}c).

\subsection*{Comparison of magnetic torques between microscopic and continuum discreptions}
To derive the specific expression of the coefficient of the magnetic torque in equation \eqref{eq7}, we simply write down the evolution equation of the angle $\theta$, defined by $\textbf{n}(\textbf{r},t) = (\cos\theta, \sin\theta)$, with respect to only the magnetic torque:
\begin{equation} \label{eq17}
\begin{pmatrix} -\sin\theta \\ \cos\theta \end{pmatrix}\frac{d \theta}{dt} = J_B\cos\theta\begin{pmatrix} \sin^2\theta \\ -\sin\theta\cos\theta \end{pmatrix},
\end{equation} 
\begin{equation} \label{eq18}
\frac{d \theta}{dt} = -J_B\cos\theta\sin\theta.
\end{equation} 
Equation \eqref{eq14} readily reads
\begin{equation} \label{eq19}
\frac{d \theta}{dt} = -2\Gamma\epsilon_1 B^2\cos\theta\sin\theta,
\end{equation} 
thus giving $J_B = 2\Gamma\epsilon_1 B^2$.

\section*{Data availability} 
All the raw data in the main text can be found in Zenodo. The link will be added after the manuscript is accepted.

\section*{Code availability} 
The custom-made codes used for the analysis of the experiment and the numerical calculation are available from the corresponding author upon request.

\section*{Acknowledgements}
We thank F. Sohrabi for technical support in bacterial culture, T. K\"{a}rki and L. Laiho for establishing protocols for bacterial culture, and C. Rigoni for building the Helmholtz coil setup. J.V.I.T acknowledges funding from ERC (803937). This work was carried out under the Academy of Finland Center of Excellence Program (2022-2029) in Life-Inspired Hybrid Materials (LIBER), project number (346112). K.B. acknowledges support from the Overseas Postdoctoral Fellowship of the Uehara Memorial Foundation.

\section*{Author contributions} 
K.B. and J.V.I.T. designed the project and wrote the manuscript. K.B. performed the experiments, the data analysis, and the theoretical analysis.

\section*{Competing interests}

The authors declare no competing interests.

\section*{Supplementary Videos}
\noindent\textbf{Video 1}. Swimming bacteria in a dilute bacterial suspension with a 10\% ferrofluid concentration under an application of the magnetic field $B = 8.0$ mT. The local trajectories of tracked bacteria for a duration of \SI{10}{\second} are displayed. Scale bar, \SI{100}{\micro\meter}.

\noindent\textbf{Video 2}. Swimming bacteria in a dilute bacterial suspension with a 10\% ferrofluid concentration under an application of the magnetic field $B = 28.2$ mT. The local trajectories of tracked bacteria for a duration of \SI{10}{\second} are displayed. Scale bar, \SI{100}{\micro\meter}.

\noindent\textbf{Video 3}. The dense bacterial suspension with a 10\% ferrofluid concentration under an application of the magnetic field $B = 28.2$ mT. (Left) The bright-field movie. (Right) The color map of the orientation field overlaid on the bright-field images. Scale bars, \SI{100}{\micro\meter}.

\noindent\textbf{Video 4}. The vorticity map with velocity streamlines obtained from Video 3. Scale bar, \SI{100}{\micro\meter}.

\noindent\textbf{Video 5}. The zoom-in of Supplementary Video 3. This video is played at 1/2 speed of real-time. Scale bar, \SI{20}{\micro\meter}.

\end{document}